# Modeling Popularity in Asynchronous Social Media Streams with Recurrent Neural Networks

**Swapnil Mishra, Marian-Andrei Rizoiu and Lexing Xie**
The Australian National University, Data 61, CSIRO, Australia
{swapnil.mishra,marian-andrei.rizoiu,lexing.xie}@anu.edu.au

**Abstract**

Understanding and predicting the popularity of online items is an important open problem in social media analysis. Considerable progress has been made recently in data-driven predictions, and in linking popularity to external promotions. However, the existing methods typically focus on a single source of external influence, whereas for many types of online content such as YouTube videos or news articles, attention is driven by multiple heterogeneous sources simultaneously – e.g. microblogs or traditional media coverage. Here, we propose RNN-MAS, a recurrent neural network for modeling asynchronous streams. It is a sequence generator that connects multiple streams of different granularity via joint inference. We show RNN-MAS not only outperforms the current state-of-the-art Youtube popularity prediction system by 17%, but also captures complex dynamics, such as seasonal trends of unseen influence. We define two new metrics: the *promotion score* quantifies the gain in popularity from one unit of promotion for a Youtube video; the *loudness level* captures the effects of a particular user tweeting about the video. We use the loudness level to compare the effects of a video being promoted by a single highly-followed user (in the top 1% most followed users) against being promoted by a group of mid-followed users. We find that results depend on the type of content being promoted: superusers are more successful in promoting Howto and Gaming videos, whereas the cohort of regular users are more influential for Activism videos. This work provides more accurate and explainable popularity predictions, as well as computational tools for content producers and marketers to allocate resources for promotion campaigns.

## 1 Introduction

Popularity refers to the total attention that a digital item receives. Understanding popularity has been an important problem since the early days of social media research. Important open questions include explaining observed popularity, predicting popularity in the future, and being able to reason about the cause and effect of intervention from individuals and groups. Insights into popularity dynamics can help content producers to better prioritize production and schedule promotions, and help content providers to allocate resources for hosting and advertising.

Successful recent models of popularity fall into two categories. The first describes individual user actions, or discrete events in continuous time (e.g. tweets) (Du et al. 2016; 2013; Mishra, Rizoiu, and Xie 2016; Shen et al. 2014; Yu et al. 2016; Zhao et al. 2015). The second is based on aggregate metrics of user actions (Cheng et al. 2014; Martin et al. 2016) or aggregated event volumes (e.g. the number of daily views) (Szabo and Huberman 2010; Pinto, Almeida, and Gonçalves 2013; Rizoiu et al. 2017; Yu, Xie, and Sanner 2015). Each of these models specialise in a distinct data type, but it is common to observe data of different types for the same online item. It is desirable to develop a model that accounts for multiple heterogeneous series.

Furthermore, many popularity models provide black-box predictions (Zhao et al. 2015; Martin et al. 2016; Mishra, Rizoiu, and Xie 2016). In practice one often demands simulations on various *what-if* scenarios, such as to quantify the effect of a unit amount of promotions, to capture seasonality or the response to outliers, to name a few. Lastly, the influence users have on popularity has been subject to constant debate in this research area. The view that one or a few influential champions can make or break a cascade (Budak, Agrawal, and El Abbadi 2011) contrasts with the view that popularity largely results from a large number of moderately influential users (Bakshy et al. 2011). It is desirable to have one model on which the future effect of different users can be comparably studied.

This work aims to explain and predict the popularity of an online item under the influence of multiple external sources, in different temporal resolutions – such as both promotion events (e.g. tweets) and volumes (e.g. number of shares per day). In particular, we propose RNN-MAS[1] (Recurrent Neural Networks for Multiple Asynchronous Streams), a flexible class of models learnable from social cascades that can describe heterogeneous information streams, explain predictions, and compare user effects for both individuals and groups. Recurrent neural network is an effective tool for sequence modeling in natural language and multiple other domains (Elman 1990; Graves 2013; Sutskever, Vinyals, and Le 2014). We link multiple recurrent neural networks by allowing them to exchange information across different asynchronous streams. This is an extension to the recent RMTPP (Du et al. 2016) for a single social event sequence. We illustrate the effectiveness of this model for predicting popularity of YouTube videos under the influence of both tweeting events and sharing volumes – improving state of the art prediction by 17%.

[1]Download data and code here: `https://git.io/vx7Tk`.

We propose several new ways to interpret and simulate popularity, and implement them for RNN-MAS. The first is a *unit promotion response* metric, measures the gain in popularity per unit of promotion. Measured at different times and promotion scales, it can describe the time-varying and nonlinear effect of online promotions. The second measure, *unseen response*, captures the effect of unobserved external influence. Since neural nets is a flexible function approximator, we show that this measure can successfully capture seasonal effects. To understand the influence of users, we compute a new metric, *loudness level*, as the log-ratio of marginal gain from users. It is used to quantify the popularity gain from powerful users and moderately influential groups of users for each video. We observe that for superusers are more effective in a minority (37%) of *Nonprofit and activism* videos, whereas superusers dominate in other video categories such as *HowTo & style and Gaming*.

The main contributions of this paper include:

- RNN-MAS, a new and flexible model that links multiple asynchronous streams for predicting online popularity.
- New measures, *unseen response* and *promotion score*, to quantify content virality and explain different factors that affect popularity.
- A method for quantifying Twitter user influence on disseminating content on YouTube, proposing a new *loudness* metric and a set of observations across diverse content types on the relative influence of superusers versus everyday users.

## 2 Background

Our proposal is at intersection of three distinct bodies of literature: Hawkes process (Hawkes 1971), Hawkes intensity process (Rizoiu et al. 2017) and Recurrent neural networks (Elman 1990; Hochreiter and Schmidhuber 1997; Graves 2013) .

**Hawkes Process.** (Hawkes and Oakes 1974) Self-exciting point processes are a special case of point processes (Daley and Vere-Jones 2008) where the arrival of an event increases the probability of observing future events. A well known self-exciting process is the Hawkes process (Hawkes 1971) where the intensity of arrival of new events – the event rate$\lambda(t)$ –, depends explicitly on all past events as:

$$\lambda(t) = \mu(t) + \sum_{t_i<t} \phi(t - t_i) \ , \tag{1}$$

where $\mu(t)$ is the rate of arrival of external events (independent of the arrival of past events). Each previous event observed at time $t_i$ affects the rate of future events through the kernel $\phi(t - t_i)$, making the process self-exciting. This kernel is known as the memory kernel. Previous literature mainly uses three families of functions as memory kernels: power-law functions $\phi^p(\tau) = (\tau + c)^{-(1+\theta)}$, used in geophysics (Helmstetter and Sornette 2002) and social networks (Crane and Sornette 2008; Zhao et al. 2015; Mishra, Rizoiu, and Xie 2016) ; exponential functions $\phi^e(\tau) = e^{-\theta\tau}$, used in financial data (Filimonov and Sornette 2013); Rayleigh functions $\phi^r(\tau) = e^{-\frac{1}{2}\theta\tau^2}$, used in epidemiology (Wallinga and Teunis 2004). All three families of kernels ensure that more recent events have higher influence on the current event rate at time $t$ – the kernels are known as *time-decaying*.

One of the limitations of employing Hawkes point processes to model popularity is the parametric design of the kernel function. For real-world data, the kernel function is unknown and it needs to be approximated in a bottom-up approach based on prior and intuition.

The model proposed in Section 3.1 can non-parametrically learn the event rate function from observed data.

**Hawkes Intensity Processes (HIP)** (Rizoiu et al. 2017) describes the volume of attention series $\xi[t]$ as a self-consistent equation:

$$\xi[t] = u[t] + \alpha s[t] + C \sum_{\tau=1}^{t} \xi[t-\tau](\tau + c)^{-(1+\theta)} \tag{2}$$

where $s[t]$ and $u[t]$ are respectively the external promotion and the unseen influence series at time $t$; $\alpha$ is the sensitivity to exogenous promotions; $\theta$ modulated the power-law memory kernel, $C$ scales with content quality and $c$ is a threshold parameter to keep the power-law kernel bounded when $\tau \simeq 0$. HIP describes the volumes of attention over fixed time intervals (e.g. daily).

Modeling popularity with HIP has two drawbacks. First, analogous to the Hawkes process, HIP also requires specifying a parametric kernel. Second, HIP models volumes of data over fixed time intervals, even when detailed information about individual external promotion events is available – e.g. which and when an user tweeted a video. The models we propose in Section 3.2 and Section 3.3 tackle these challenges via joint inference of the volume series and point process series.

**Recurrent Neural Networks (RNN)** (Elman 1990) are common sequence models where the same feed forward structure is replicated at each time step. They have additional connections from the output of previous the time step to the input of the current time step – therefore creating a recurrent structure. Their hidden state vector $\mathbf{h_t}$ can be defined recursively as:

$$\mathbf{h_t} = f(\mathbf{x_t}, \mathbf{h_{t-1}})$$

where $f$ is the feed forward network, $\mathbf{x_t}$ is the current input, $\mathbf{h_{t-1}}$ is the output from previous time step. Long short-term memory (LSTM) (Hochreiter and Schmidhuber 1997; Graves 2013) units are essentially recurrent networks with additional gated structure, defined as:

$$\begin{aligned}
\mathbf{i}_t &= \sigma(\mathbf{W}_i\mathbf{x}_t + \mathbf{U}_i\mathbf{h}_{t-1} + \mathbf{V}_i\mathbf{c}_{t-1} + \mathbf{b}_i) \\
\mathbf{f}_t &= \sigma(\mathbf{W}_f\mathbf{x}_t + \mathbf{U}_f\mathbf{h}_{t-1} + \mathbf{V}_f\mathbf{c}_{t-1} + \mathbf{b}_f) \\
\mathbf{c}_t &= \mathbf{f}_t\mathbf{c}_{t-1} \odot \tanh(\mathbf{W}_c\mathbf{x}_t + \mathbf{U}_c\mathbf{h}_{t-1} + \mathbf{b}_c) \\
\mathbf{o}_t &= \sigma(\mathbf{W}_o\mathbf{x}_t + \mathbf{U}_o\mathbf{h}_{t-1} + \mathbf{V}_o\mathbf{c}_t + \mathbf{b}_o) \\
\mathbf{h}_t &= \mathbf{o}_t \odot \tanh(\mathbf{c}_t)
\end{aligned} \tag{3}$$

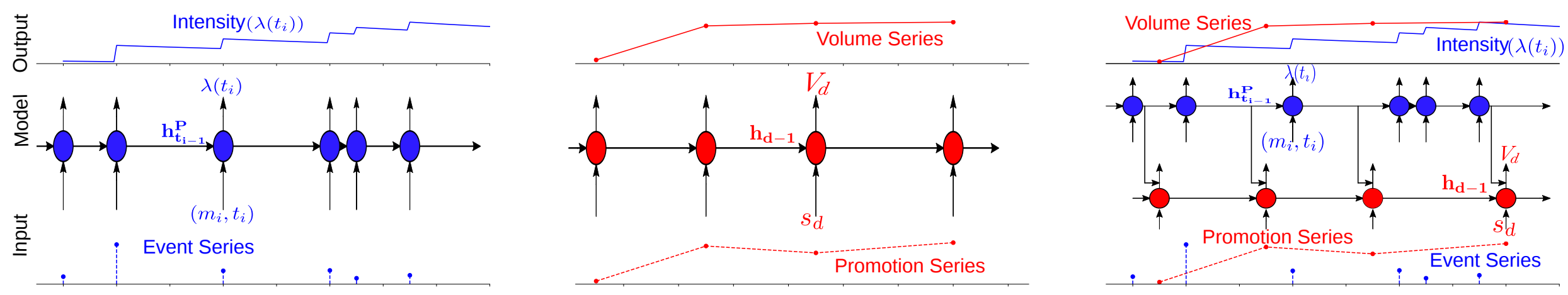

(a) (b) (c)

Figure 1: Recurrent models for events and volumes. (a) RPP, (b) VolRNN-S/TS and (c) RNN-MAS. Top row: output series; middle: block diagram for recurrent network, each circle refers to an LSTM unit; bottom: input series.

where $\mathbf{x}_t$ is input at time $t$, $\sigma$ is the logistic sigmoid function and $\odot$ denotes element-wise multiplication. The notations $\mathbf{i}_t$, $\mathbf{f}_t$, $\mathbf{c}_t$, $\mathbf{o}_t$ and $\mathbf{h}_t$ stand for the input, forget, cell-state, output and hidden state at time $t$. We use the following shorthand notation for the LSTM set of equations in Equation 3:

$$(\mathbf{h}_t, \mathbf{c}_t) = LSTM(\mathbf{x}_t, \mathbf{h}_{t-1}, \mathbf{c}_{t-1}) \tag{4}$$

LSTM and its variants have been successfully used for modeling time series and predicting sequence, due to their ability to capture the effects of past data in their hidden state (Hochreiter and Schmidhuber 1997; Graves 2013; Chung et al. 2014). We build on this intuition to construct based on LSTM the models proposed in Section 3.

## 3 RNN for Volume and Point Processes

In the following section we first introduce a recurrent model for modeling event series in Section 3.1, followed by a recurrent model for a volume series in Section 3.2. Finally we present a joint model for predicting attention volumes in Section 3.3. Our notations for all the model are sumarized in Table 1.

### 3.1 Recurrent Point Process (RPP)

We present a model to frame events data as point processes in Figure 1a, where the influence of history on future events is learned as parameters of LSTM unit.

The equation for event rate at time $t$ for the model is written as:

$$\lambda(t) = \exp\left(W_o^P \mathbf{h_t^P} + \alpha\,(t - t_i)\right) \tag{5}$$

such that $\mathbf{h_t}$ is :

$$\mathbf{h_t^P} = LSTM\left((m_i, t_i), \mathbf{h^P_{t_i}}, \mathbf{c_{t_i}}\right) \tag{6}$$

where $t_i$ is the time for last event before time $t$, $m_i$ is the magnitude of last event. In point process literature magnitude of an event is generally referred to as mark of an event, where mark represent attributes of the event, e.g. magnitude of the earthquake, or the type of a trade - bond or stock. In our work for modeling tweets we consider mark as the magnitude (#followers) of user tweeting. $\mathbf{h_{t_i}}$ and $\mathbf{c_{t_i}}$ denotes hidden and cell state respectively at time $t_i$, and $\alpha$ is a scalar that modulates effect of current time on intensity.

Our model is similar to the one proposed by (Du et al. 2016). In their work they assume discrete values of marks as they model different event types rather than magnitude.

Table 1: Summary of notations for RNN models

| | Symbol | Quantity |
|---|---|---|
| Event Data | $e_i = (m_i, t_i)$ | $i^{th}$ event at time $t_i$ of magnitude $m_i$ |
| | $\lambda(t_i)$ | Intensity at $i^{th}$ event |
| Volume Data | $s_d$ | Exogenous stimulus at fixed interval d |
| | $V_d$ | Attention Volume at fixed interval d |

We generalize their model to work for continuous marks by modeling mark generation separately from history, as done in (Rizoiu et al. 2017; Mishra, Rizoiu, and Xie 2016).

### 3.2 Volume RNN

RNN can also model volume streams. HIP can be seen as generalization of autoregressive models, with a history size as long as the sequence itself. Intuitively in HIP, the current value of volume intensity can be seen as a convolution of a memory kernel with its past values. Hence we formulate a LSTM model in Figure 1b, for predicting attention volume series as follows, using Equation (4):

$$V_d = W_o^{Vol} \mathbf{h_d} \tag{7}$$

such that,

$$\mathbf{h_d} = LSTM(\mathbf{s_d}, \mathbf{h_{d-1}}, \mathbf{c_{d-1}}) \tag{8}$$

where $\mathbf{V}_d$ is the attention volume at time $d$ , $\mathbf{h}_d$ is the hidden state time $d$ and input to the system is $\mathbf{s}_d$ at time $d$, where $s_d$ is the exogenous stimuli for the attention volume at time $d$. For volume RNN model we have time points as discrete values, spread over fixed time intervals (e.g. daily, hourly).

### 3.3 RNN-MAS for asynchronous streams

In real world scenarios, we have multiple streams of data promoting a single attention series. For example, being tweeted and being shared both lead to more views on a YouTube video. In this section we present a combined model to take into account the volume promotions and individual events for predicting an attention volume series.

RNN-MAS model is shown in the Figure 1c. It has two components: a volume RNN model as shown in Figure 1b predicting the desired attention volume series at fixed time intervals by taking into account the promotional volume series; a second component as RPP model as shown in Figure 1a, responsible for modeling the individual promotion series. Intuitively we would like RPP to modulate the response of volume RNN, to make better predictions for attention series. We achieve it by combining the hidden states for individual models just before making the prediction for attention series.

We combine individual models as follows:

$$V_d = W_o^{MAS} \left[\mathbf{h_d}, \mathbf{h_d^P}\right] \tag{9}$$

where, $\mathbf{h_d}$ and $\mathbf{h_d^P}$ are the hidden state for volume RNN and RPP on day d, calculated as per Equation 8 and Equation 6 respectively, shown with arrows coming from RPP model towards the volume RNN model at regular intervals in Figure 1c.

We note another plausible setup for joint modeling is attention volumes affecting the event rate of RPP. As our main concern is predicting popularity under promotion we do not utilize this setting.

### 3.4 Model Learning

**RPP:** For fitting our model to a sequence of events $S = \{e_i\}$ where, $e_i$ is the $i^{th}$ event such that $m_i$ and $t_i$ stands for magnitude and time of the $i^{th}$ event, we maximize the log-likelihood of the observed sequence. Using Equation 5 and Equation 6, we can write log-likelihood for sequence as:

$$\begin{aligned} LL\left(S\right) &= \sum_{i=1}^{|S|} log\left(\lambda\left(t_i\right)\right) - \int_0^{t_{|S|}} \lambda\left(\tau\right) d\tau \\ \implies LL\left(S\right) &= \sum_{i=1}^{|S|} \Big[ W_o^P \mathbf{h_{t_i}^P} + c\tau_i + \frac{1}{c}\exp\left(W_o^P \mathbf{h_{t_i}^P}\right) \\ &\qquad - \frac{1}{c}\exp\left(W_o^P \mathbf{h_{t_i}^P} + c\tau_i\right)\Big] \end{aligned} \tag{10}$$

For predicting the time for the next event in sequence we can utilize the relationship between conditional density function of time, $f\left(t\right)$, and conditional intensity function (event rate), $\lambda\left(t\right)$ (Daley and Vere-Jones 2008):

$$f\left(t\right) = \lambda\left(t\right) \exp\left(-\int_{t_i}^{t} \lambda\left(\tau\right) d\tau\right) \tag{11}$$

In order to predict the next time step,$\hat{t}_{i+1}$, where $t_i$ is event time for the last observed event, we take the expectation in Equation 11 as follows:

$$\implies \hat{t}_{i+1} = \int_{t_i}^{\infty} t \cdot f\left(t\right) dt \tag{12}$$

**Volume RNN:** For training volume models, we use *RMSE* as the loss function and each fixed interval data is considered as separate time steps of the given variables. This can be seen as modeling a time-series regression with LSTM.

**RNN-MAS:** For training the joint model, we first train both RPP and volume RNN models independently. After fitting both models, we calculate the hidden state from RPP model for day d using Equation 6 and hidden state for volume RNN at day d is calculated using Equation 8. Now the combined model is trained using *RMSE* for the attention volume series.

For all models, we jointly learn network parameters over 500 random sequences from `Active+tweets Dataset` described in Section 5. Now for individual fittings we use the jointly learned parameters as the initialization point. We observe faster convergence and better prediction results when compared to running with random initializations. For more details, see supplement (Mishra, Rizoiu, and Xie 2018).

## 4 Popularity Metrics

RNN-MAS describes popularity under influence of heterogeneous streams of shares and tweets. We propose two metrics based on RNN-MAS to quantify average response to unit promotion, and the relative influence among users of different *fame*. We also utilize the model to estimate a response series to unseen influence for a video.

### Simulation

We can use our learned volume prediction models to simulate a series of views $V_d^{P(\cdot)}$ from $d = 1$ to $d = T$ days, where $V_d^{P(\cdot)}$ stands for views on $d^{th}$ day under a promotion function $P(\cdot)$. For computing views we evolve hidden states as per the Equation 6 and Equation 8 for our models, where the input parameter is defined by the value of function $P(\cdot)$ at various time steps. At the end of each time step, views are computed as per Equation 9 and Equation 7 for RNN-MAS and Volume RNN respectively. We define, $\nu\left(P(\cdot)\right)$, the cumulative views generated by promotion function $P(\cdot)$ from $d = 1$ to $d = T$ days as:

$$\nu\left(P(\cdot)\right) = \sum_{d=1}^{T} V_d^{P(\cdot)} \tag{13}$$

For all our simulations we choose $T = 10,000$ days, as this represents ≈ 27 years, longer than lifetime of any YouTube video.

### 4.1 Response to unseen influence

Despite taking into account multiple sources of external influence, there are influence signals that our models, RNN-MAS and volume RNN, do not capture explicitly. Examples include seasonality, or discussions in forums that are known to be an important factor for *gaming* videos — widely promoted on forum `www.minecraftforum.net`.

For understanding this response to unseen influence we set the promotion function $P(\cdot) = 0$ in our simulation setup described in Section 4 and obtain a series of $V_d^{P(\cdot)=0}$.

The response series to unseen influence here generalizes previous definitions. In HIP (Rizoiu et al. 2017) external influence $u[t]$, is assumed to be an initial impulse plus a constant (Equation 5 in Section 2.4). In this work the shape of response series to latent promotions is unconstrained.

In Section 6.1 we present various case studies to illustrate how the response series to unseen influence for our models is able to capture richer temporal trends, such as seasonality.

### 4.2 Response to unit promotion.

In linear time variant (LTI) systems such as HIP, the total gain per unit promotion, $\nu$,is well defined. It's calculated by computing the impulse response (Rizoiu and Xie 2017) of the system. In our models, RNN-MAS and volume RNN, the notion of impulse response does not readily apply as it's a non-linear time variant system.

We compute the average response to unit impulse in three steps: compute response of our models to $p$ units of promotion to volume RNN at $d = 0$ days; subtract response to unseen influences, calculated as per Section 4.1; normalize it by $p$. Response to $p$ units of promotion is calculated using function $P(d) = p\mathbb{1}[d]$ in simulation setup of Section 4, where $\mathbb{1}[d]$ takes a value of 1 at $d = 0$, and 0 otherwise. The average response to unit promotion from promotion $P(d) = p\mathbb{1}[d]$, $\varrho\,(p)$ named as promotion score, is calculated as:

$$\varrho\,(p) = \frac{\nu(p\mathbb{1}[d]) - \nu(0)}{p} \tag{14}$$

In this definition, promotion score can be negative, capturing cases where being promoted decreases popularity (e.g. spam or paid tweets). Also note that $\varrho(p)$ is a function of p, i.e. for different simulation parameter p, the average response to unit promotion is different. This allows us to describe phenomena of diminishing returns in may real-world marketing scenarios (Jones 1990). We examine the effect of p in Section 6.2 and found that $\varrho(p)$ tend to decrease as p grows.

### 4.3 Loudness level

Using simulation metrics similar to $\vartheta$ and $\varrho(p)$, we can measure the influence of an individual or a group of users by the responses generated by their tweeting events.

Loudness level is a relative measure of total influence for one or a group of users. It is measured in decibels (dB). It is computed as log-ratio of the response of the target group $\psi(S)$ versus that of a comparison value $\psi_0$.

We compute $\psi(S)$ in two steps: calculate $\nu(P(\cdot))$ such that $P(\cdot) = \{s_i\}$ in simulation setup of Section 4, where $\{s_i\}$ is the set of tweets generated by user group $S$; subtract response to unseen influences, calculated as per Section 4.1. It can be written as:

$$\begin{aligned} &\psi(S) = \nu(s_i) - \nu(0) \\ \implies\ &\psi_{dB}(S) = \log_{10}\left(\frac{\psi(S)}{\psi_0}\right) dB \end{aligned} \tag{15}$$

where $\psi_0$, the comparison value, is set to 1 for experiments in Section 6.3.

As loudness level accounts for all activity generated by a user group, it can be used to compare the relative effects of promotion between users of different *fame* in Twitter.

## 5 Predictive Evaluation

In this section, we evaluate the performances of predicting the attention that videos receive in the future using RNN-MAS and its variants, against a number of baselines.

### Overview of methods

**Event prediction.** We compare several approaches for predicting either the next event time in a series of events, or the likelihood of a series of future (holdout) events:

- **RMTPP (Du et al. 2016):** State of the art system for modeling event data with recurrent point processes.
- **RPP:** Our model detailed in Section 3.1, modeling event data with LSTM.
- **PL (Mishra, Rizoiu, and Xie 2016):** State of the art model for events data utilizing Hawkes processes.
- **Exp:** variant of PL, that uses an exponential kernel.
- **Seismic (Zhao et al. 2015):** Seminal model based on self exciting processes for predicting event data in social networks.

**Volume prediction.** We use the following approaches to predict the series of volume of attention that a video receives during the next time-frames:

- **HIP (Rizoiu et al. 2017):** State of the art system for predicting views of video by using daily shares as promotions, outperforms linear regression baselines (Pinto, Almeida, and Gonçalves 2013) and (Szabo and Huberman 2010).
- **VoRNN-S:** Our model detailed in Section 3.2 which like HIP, uses only daily shares for prediction.
- **VoRNN-TS:** A version of our model detailed in Section 3.2, using both daily shares and tweets as promotions, where tweets are aggregated daily to make it synchronous to views and shares series.
- **RNN-MAS:** Our dual RNN model described in Section 3.3 that combines daily shares with the event series of tweets, modeled as point process by RPP, to make predictions.

### Dataset

We use three different datasets in our experiments.

**RMTPP-Syn:** synthetic data simulated with same parameters as (Du et al. 2016). It has 100,000 events from an unmarked Hawkes process (Hawkes and Oakes 1974). The conditional intensity for simulated data is given by $\lambda\,(t) = \lambda_o + \alpha \sum_{t_i<t} exp\left(-\frac{t-t_i}{\sigma}\right)$ where $\lambda_o = 0.2$, $\alpha = 0.8$ and $\sigma = 1.0$. We use $90\%$ of events for training and rest for testing.

**Tweet1MO:** a real world tweets dataset released by (Zhao et al. 2015), containing all tweets between October 7 to November 7, 2011.

**Active+tweets:** The YouTube videos dataset released by (Rizoiu et al. 2017) containing 13,738 videos. All videos in this dataset were uploaded between 2014-05-29 to 2014-08-09 and have at least 100 views, 100 shares and 100 tweets. We augment each video with corresponding individual tweets data, emitted during same period. There are in total 30.2 million tweets in the dataset. On average each video is tweeted 2151 times, median video was tweeted 327 times.

### 5.1 Event Data Results

Table 2 lists two prediction tasks, next event time and holdout likelihood, against competing methods for the task. We present results for next event time prediction on RMTPP-Syn as Tweet1MO has continuous marks which RMTPP can not model. Seismic can not predict next event time hence we compare performance for hold-out likelihood on Tweet1MO.

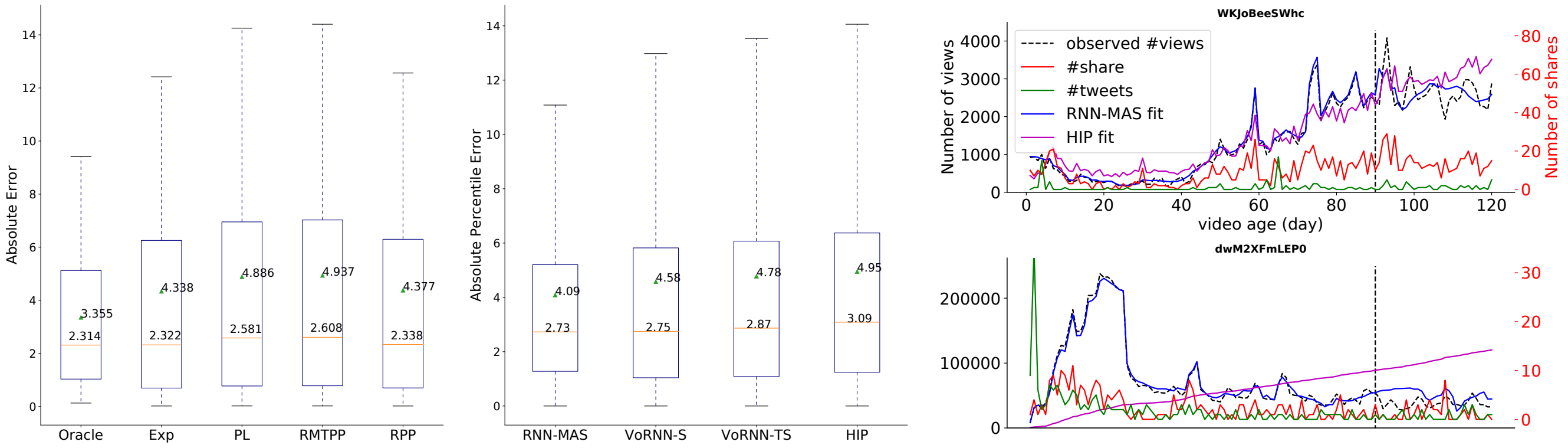

(a) (b) (c)

Figure 2: (a) Error in estimated next event time for various models on RMTPP-Syn. (b) Absolute Percentile Error(APE) for different approaches on *Active* dataset. All approaches use shares as promotion, VoRNN-TS and RNN-MAS use extra information about tweets. (c) two sample fittings where error(APE) for RNN-MAS is 2.01 and 0.13 and error for HIP is 2.3 and 5.1 for videos 'WKJoBeeSWhc' and 'dwM2XFmLEP0' respectively.

Table 2: Summary of methods and prediction tasks for event data modelling.

| Method | Prediction Type | |
|---|---|---|
| | Next Event (RMTPP-Syn) | hold-out likelihood (Tweet1MO) |
| RMTPP | ✓ | ✗ |
| Seismic | ✗ | ✓ |
| PL | ✓ | ✓ |
| EXP | ✓ | ✓ |
| RPP | ✓ | ✓ |

We make prediction for next event time for RMTPP, PL, Exp and RPP using Equation 12. Figure 2a reports error in prediction as absolute error. We report error for an oracle process that is set to the ground truth parameters generating data. Error for oracle process is difference between the expected values and simulated values. We observe that RPP has lower mean and median error than RMTPP. RPP shows similar mean and median as EXP, indicating RPP is flexible enough to capture influence of past events on future.

We compare the negative log-likelihood on hold-out data for RPP and alternatives, lower is better. We vary the fraction of observed events to report results on Tweet1MO in Table 3.

We observe RPP performs better than PL and Seismic for all observed fractions. More importantly performance of RPP increases consistently with increase in observed fraction, indicating better generalization towards data.

Table 3: Results showing mean(median) of negative log-likelihood per event for holdout set for Tweet1MO after observing a part of the cascade.

| Observed Fraction | PL | Seismic | RPP |
|---|---|---|---|
| 0.40 | 8.54(7.87) | 7.62(6.98) | 7.46(6.47) |
| 0.50 | 8.38(7.96) | 7.44(6.44) | 7.26(6.39) |
| 0.80 | 9.22(9.29) | 8.02(7.3) | 6.97(6.14) |

## 5.2 Volume Prediction

**Setup.** We evaluate our volume prediction approaches in the same a temporal holdout setup employed by (Rizoiu et al. 2017), on the Active+tweets dataset. Four data streams are available for each video in this dataset: the views volume series, the shares volume series, the tweets volume series and the tweet event data series. Each of the models listed in Table 4 inputs one or several promotion streams (shares and tweets) and predicts the views series stream. First, we train each model using the streams data in the first 90 days, for each video. Next, we forecast the views series during the next 30 days (from day 91 to 120) assuming known the promotions streams. Finally, we compute the gained popularity by summing up the views series during the test period. We evaluate performances using the Absolute Percentile Error (APE), computed as in (Rizoiu et al. 2017). Note that out of the approaches in Table 4, RNN-MAS is the only model that can leverage the tweet event data stream for forecasting the futures views series.

**Results.** Figure 2b shows the results for `Active+tweets Dataset`. We observe that VoRNN-TS, outperforms HIP by 7% and 3% for median and mean error, respectively. RNN-MAS performs best with an improvement of 11% and 17% for median and mean error respectively over HIP. We observe an increase in performance of RNN-MAS when compared with VoRNN-S whereas for VoRNN-TS performance decreases, despite both models utilizing additional tweets data over shares in VoRNN-S. We are modeling this extra information in VoRNN-TS as volume of tweets wherein in RNN-MAS we utilize the timing (and magnitude) of individual tweets. Results show that timing of events appears to contain information useful for predicting views, hence necessitating a model to jointly model event and volume data.

[2]Not used as performance decreases

Table 4: Summary of methods used in the volume evaluation, and type of promotion data that each approach can use.

| Method | Type of Promotion | | |
|---|---|---|---|
| | Shares (daily) | Tweets (daily) | Tweets (events) |
| HIP | ✓ | ✓[2] | ✗ |
| VoRNN-S | ✓ | ✗ | ✗ |
| VoRNN-TS | ✓ | ✓ | ✗ |
| RNN-MAS | ✓ | ✓[3] | ✓ |

Figure 2c shows two sample fittings, for videos *WKJoBeeSWhc* and *dwM2XFmLEP0* where error(APE) for RNN-MAS is 2.01 and 0.13 and error for HIP is 2.3 and 5.1 respectively. The vertical dashed line divides data into training and testing set of 90 and 30 days respectively. The dashed line shows the real view series where blues line shows the fit by RNN-MAS and magenta shows the fit for HIP. For video *WKJoBeeSWhc* you could see two different phases, one before 90 days where view series (shown with dashed black line) is showing an inherent upward trend and one after 90 days where this inherent tendency has died. Fitting for RNN-MAS (shown in blue) is able to capture this change wherein HIP (shown in magenta) fails. For video *dwM2XFmLEP0* performance of HIP is even worse as it is not able to model the view series in accordance with shares (shown in red) and RNN-MAS ably models the response. One explanation is that RNN-MAS captures latent influence better, as shown in Section 6.1.

# 6 Explaining Popularity

In this section, we exemplify how the response series to unseen influence, and how the metrics *promotion score* and *loudness level* described in Section 4 can be used to analyze the popularity of online videos.

## 6.1 Understanding unseen influence

We present case studies on real and simulated data to illustrate the complex dynamics captured by the response series to unseen influence for our models when compared to HIP.

**Real data fittings.** We show sample fittings for two videos *WKJoBeeSWhc* and *1hTDORFb5SE* in the top row of Figure 3a and Figure 3b respectively. The bottom row of both figures shows the response series to unseen influence for RNN-MAS and HIP. Visibly, RNN-MAS can capture complex dynamics that HIP cannot explain: video *WKJoBeeSWhc* (Figure 3a) exhibits a delayed response, while *1hTDORFb5SE* (Figure 3b) features a seasonal trend. Part of the fitting performance gain of RNN-MAS can be attributed to its modeling of the response to unseen influence.

**Synthetic data.** We further investigate on synthetic data the ability of volume RNN models and HIP to uncover seasonal trends. We simulate two sources of promotion for 120 days: the first series has a weekly cyclic component, with a maximum amplitude of 24 units of promotion (shown in the bottom row of Figure 4a); the second series contains random promotions with the same maximum amplitude (shown in the middle row of Figure 4a). Using these two promotion series, we simulate the view series using HIP (shown in the top row of Figure 4a). Next, we use the fitting procedure described in Section 5.2 to train the HIP and the VoRNN-S models on the simulated view series, using only the random promotion as the external stimulus series. More details on the simulation can be found in the online supplement (Mishra, Rizoiu, and Xie 2018).

The top row of Figure 4b shows the fitting results for both models. Visibly, the series fitted by VoRNN-S (blue line) follows the simulated views (dashed black lines) more closely than the series fitted by HIP (magenta line). The bottom row of Figure 4b shows the response to unseen promotions, as estimated by VoRNN-S (in blue) and HIP (magenta). We see that VoRNN-S uncovers a cyclic response compatible in phase with the unobserved promotion series. The response of HIP shows a sharp drop followed by a near-zero constant response, due to its simplistic external influence modeling.

## 6.2 Average unit promotion score

HIP was used in (Rizoiu and Xie 2017) to quantify the total amount of attention generated by a single unit of promotion – dubbed the *virality score* $\nu$ – for an online promotion application. In Figure 3c, we compare the promotion score $\varrho(p)$ of our models against $\nu$ for all the videos in the `Active+tweets Dataset`. The x-axis shows the promotion score $\varrho(p)$ with $p = 5$, split into 20 percentile bins. The y-axis shows $\nu$ in percentiles, summarized using boxplots for each bin. Overall, we observe a strong agreement between $\varrho(p)$ and $\nu$ – the boxplots are placed around the main diagonal – signifying that RNN-MAS can capture online promotability as well as HIP does. Furthermore, we observe a decrease in the value of $\varrho(p)$ as $p$ increases: the median values are $\varrho(1) = 263$, $\varrho(5) = 260$, $\varrho(10) = 259$. This suggests that, unlike HIP, our models can capture the complex phenomenon of diminishing returns in promotion (Jones 1990).

## 6.3 User Influence

**Setup.** Bakshy et al (Bakshy et al. 2011) showed that for promoting content on Twitter, the most reliable and cost-effective method is to utilize a group of individuals who have average or even less number of followers. We evaluate the influence of a super user against a cohort of small users. Figure 5a shows the rank versus the number of followers in `Active+tweets Dataset`. We define a *super* user as having 21,368 followers, or top 1% of all users; a *small* user as the median user with 120 followers. The cohort has 178 *small* users, as the sum total of their followers is comparable to a super user. For each video and each user group we generate a series of tweets from the trained RPP model of the video. This series of tweets act as promotions by each user group. The effect for tweet series promotion is captured by metric $\psi_{dB}$ for each group as described in Section 4.3. Setup details are described in the supplement (Mishra, Rizoiu, and Xie 2018).

[3] Not used, theoretically it can

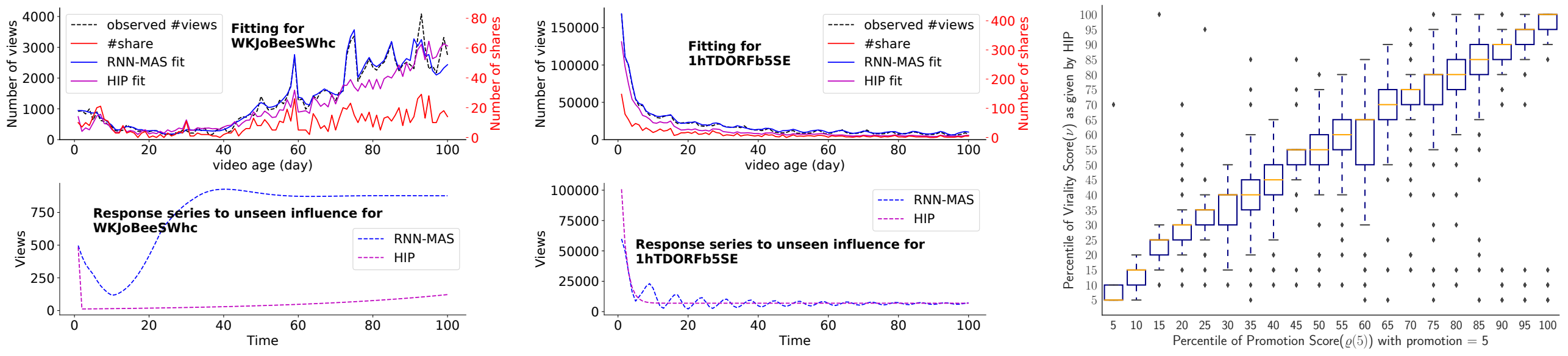

(a) (b) (c)

Figure 3: **(a)(b)** Fittings for the attention series (top row) and the response to unseen influence (bottom row) using RNN-MAS and HIP, for two sample videos *WKJoBeeSWhc* (a) and *1hTDORFb5SE* (b). **(c)** Comparison of HIP's virality score $\nu$ (x-axis) against the promotion response of RNN-MAS $\varrho(p = 5)$ (y-axis).

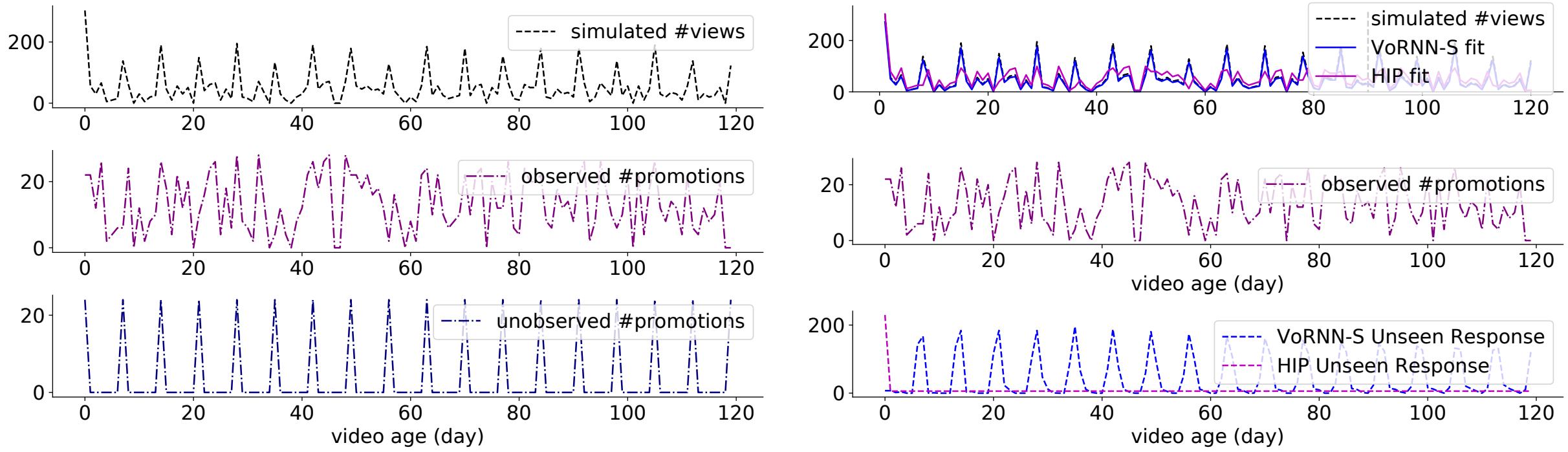

(a) (b)

Figure 4: **(a)** Simulating a views series with known unobserved influence. A view series (top row) simulated by HIP with two sources of external stimuli: a series of random promotions (middle row) and a cyclic promotion series with weekly spikes (bottom row). **(b)** Retrieving the reaction to unknown influence through fitting. The simulated views series in (a, top row) is fitted using VoRNN-S and HIP (shown in the top row) using only the observed influence series (middle row). The reaction to unobserved influence (bottom row) is shown for HIP (magenta) and VoRNN-S (blue).

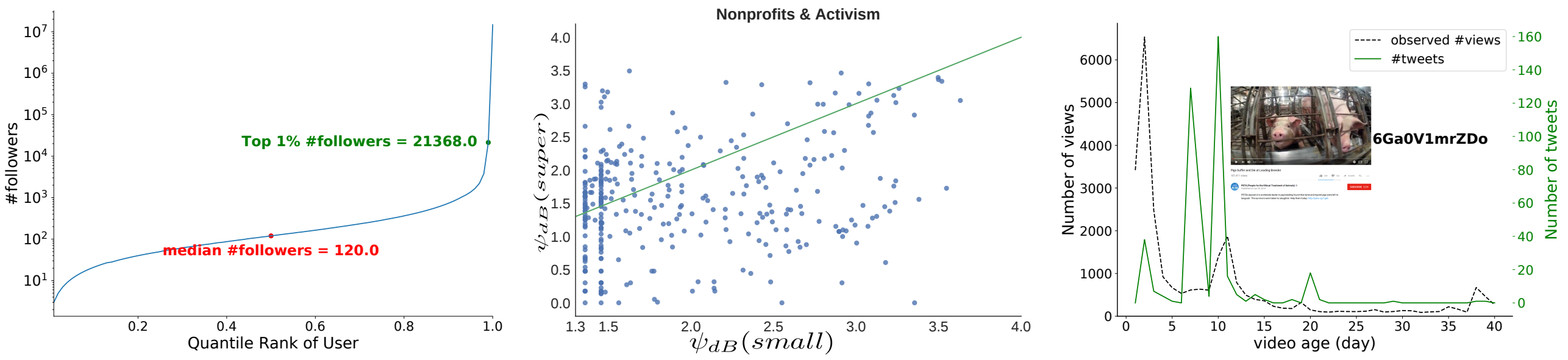

(a) (b) (c)

Figure 5: (a) Quantile rank versus number of followers of an user on `Active+tweets Dataset`. The median user has 120 followers whereas a user with 21368 followers is in top 1%. (b) $\psi_{db}(super)$ versus $\psi_{db}(small)$ scatter plot for *Nonprofits & Activism* videos. For 37% of the videos, super users have greater influence than the small user cohort. (c) View and tweet series for video *6Ga0V1mrZDo*,with $\psi_{db}(super)$ = 3.3 and $\psi_{db}(small)$ = 1.3. See Sec 6.3 for discussions.

**Results.** Figure 5b shows the value of $\psi_{dB}$ for super user against a cohort of small users on all videos in *Nonprofits & Activism* category. Plots for all categories are in the supplement (Mishra, Rizoiu, and Xie 2018). We observe type of content greatly affects the promoting capabilities of the super user. For example 37% of *Nonprofits & Activism* videos (above the green line) have $\psi_{dB}$ of super user greater than $\psi_{dB}$ of cohort of small users. Whereas for categories *HowTo & Style* and *Gaming* more than 50% of videos have a greater $\psi_{dB}$ for super user when compared to $\psi_{dB}$ for a cohort of small users.

We present an example video (*6Ga0V1mrZDo*) from *Nonprofits & Activism* category in Figure 5c, where $\psi_{dB}(super)$ is 3.3 and $\psi_{dB}(small)$ is 1.3. The video here is uploaded by *PETA*, talking about deaths of pigs in slaughter houses. In Figure 5c, not all peaks in tweets corresponds towards the increase in views of the video. On examining the data we find that peaks where views and tweets correlates are the points where a super user is part of the tweets. In first peak around day 3, there is a user with 494851 followers. In the third peak around day 12, there is a user with 23561 followers. At two other tweeting peaks, there are cohorts of small users (all with less than 4000 followers) tweeting about the video, but no view count pikes. Presence of super users around peaks in view series corroborates the loudness estimate for this video.

## 7 Related Work

**Modeling and predicting popularity using generative models.** Popularity modeling (Crane and Sornette 2008; Yu et al. 2016) and prediction (Shen et al. 2014; Zhao et al. 2015; Mishra, Rizoiu, and Xie 2016) has been a particularly fertile field for point-process based generative models. In their seminal work, Crane and Sornette (Crane and Sornette 2008) showed how a Hawkes point-process can account for popularity bursts and decays. Afterwards, more sophisticated models have been proposed to model and simulate popularity in microblogs (Yu et al. 2016). These approaches successfully account for the social phenomena which modulate online diffusion: the "rich-get-richer" phenomenon and social contagion. Certain models can output an estimate for the total size of a retweet cascade. Shen *et al.* (Shen et al. 2014) employed reinforced Poisson processes, modeling three phenomena: fitness of an item, a temporal relaxation function and a reinforcement mechanism. Zhao *et al.* (Zhao et al. 2015) proposed SEISMIC, which employs a double stochastic process, one accounting for infectiousness and the other one for the arrival time of events. The model by (Mishra, Rizoiu, and Xie 2016) uses both hawkes process and feature driven methods to predict total tweets for a cascade. HIP (Rizoiu et al. 2017; Rizoiu and Xie 2017), current state of the art, uses volume point process to directly model views of a YouTube video and then extends it for promotion and marketing. (Wang et al. 2017a) recently proposed a method to link macro and micro events of a single time series to do macro prediction, they aggregate the micro time series to get the macro time series.

**RNN for time series.** Recurrent Neural Networks (Elman 1990) in particular its variants LSTM (Hochreiter and Schmidhuber 1997; Graves 2013) and GRU (Chung et al. 2014) are the basic building block of our model. They are basically a sequence learning model that generates the next output in a sequence depending upon its current input and previous state. They have shown state of the art results in sequence prediction/learning task in fields text (Sutskever, Vinyals, and Le 2014), image (Vinyals et al. 2015), video (Jain et al. 2016) and time-series data (Chandra and Zhang 2012). Recently people have also tried to model point processes (Du et al. 2016; Xiao et al. 2017; Wang et al. 2017b) with RNNs. Recent work by (Cao et al. 2017; Li et al. 2017) used RNNs for modeling end to end prediction of information cascades in presence of a single source of events.

The main difference in our model with any existing work is the use of two different streams of data, one is volume and other is event series, about a single event. Although HIP (Rizoiu et al. 2017; Rizoiu and Xie 2017) uses two series but instead of using event series as it is they aggregate it and report worse performance then using only the macro series. Recent work by (Wang et al. 2017a) and (Xiao et al. 2017) combines asynchronous streams, but they aggregate exogenous individual events at fixed intervals rather than considering a separate exogenous volume stream.

**Influence of users of different *fame*.** (Cha et al. 2010; Romero et al. 2011) showed in their work that influence differs for users with topics and high number of followers is not related to high influence in a network. Work by (Bakshy et al. 2011) has shown that a cohort of small users are as powerful as a big user in case of information diffusion on Twitter whereas work by (Budak, Agrawal, and El Abbadi 2011) has shown that in cases for limiting misinformation in network targeting big users is more beneficial.

To the best of our knowledge there exists no prior work in estimating a user influence across network boundaries. Our proposed metric, loudness level, along with joint model, RNN-MAS, helps us to estimate the influence of a Twitter user in spreading content on YouTube.

## 8 Conclusion

This paper proposed RNN-MAS, a model for predicting popularity under the influence of multiple heterogeneous asynchronous streams – namely tweets and volumes of shares for a YouTube video. With this model, we demonstrated superior performance on forecasting popularity on a large-scale YouTube video collection. We further design two new measures, to explain the viral potential of videos, another to uncover latent influences including seasonal trends. One important application of such model is to compare effects of different kinds of promotions. We propose a new metric, dubbed *loudness*, to quantify the relative effective-ness of user groups. We show that superusers and grassroot are effective in different content types.

While neural network are expressive function approximators that can learn nonlinear relations between input and output series, RNN-MAS does not model temporal dynamics drifts or finite population effects. We hope to extend this work to allow different items to share parts of the model, and eventually make better predictions as early as possible in a digital item's lifetime.

**Acknowledgments** This material is based on research sponsored by the Air Force Research Laboratory, under agreement number FA2386-15-1-4018. We thank the National Computational Infrastructure (NCI) for providing computational resources, supported by the Australian Government.